# Groups of Highly Cited Publications: Stability in Content with Citation Window Length

Nadine Rons[1]

[1] Nadine.Rons@vub.ac.be
Vrije Universiteit Brussel, Research Coordination Unit and Centre for R&D Monitoring (ECOOM), Pleinlaan 2, B-1050 Brussels (Belgium)

**Introduction**

The growing focus in research policy worldwide on top scientists makes it increasingly important to define adequate supporting measures to help identify excellent scientists. Highly cited publications have since long been associated to research excellence. At the same time, the analysis of the high-end of citation distributions still is a challenging topic in evaluative bibliometrics. Evaluations typically require indicators that generate sufficiently stable results when applied to recent publication records of limited size. Highly cited publications have been identified using two techniques in particular: pre-set percentiles, and the parameter free Characteristic Scores and Scales (CSS) (Glänzel & Schubert, 1988). The stability required in assessments of relatively small publication records, concerns size as well as content of groups of highly cited publications. Influencing factors include domain delineation and citation window length. Stability in size is evident for the pre-set percentiles, and has been demonstrated for the CSS-methodology beyond an initial citation period of about three years (Glänzel, 2007). Stability in content is less straightforward, considering for instance that more highly cited publications can have a later citation peak, as observed by Abt (1981) for astronomical papers. This paper investigates the stability in content of groups of highly cited publications, i.e. the extent to which individual publications enter and leave the group as the citation window is enlarged.

**Data and methodology**

*Database, document type and time frame*
The bibliometric data were obtained from the online Web of Science. This study focuses on articles as primary vehicles for new research results (reviews, less frequent and typically more highly cited, are in itself related to high esteem). Results were calculated for articles published in 2004. Citations were collected up to 7 years after publication, until 2011.

*Aggregation level*
The group of top-cited publications is highly dependent on the level of aggregation at which these are determined (Zitt, Ramanana-Rahary & Bassecoulard, 2005). In this paper, the reference sets within which highly cited publications are identified are the partition cells formed by the structure of overlapping subject categories (Rons, 2012). These partition cells are an aggregation level of intermediate size, situated between journals and entire subject categories, proposed particularly for usage at the level of individual scientists.

*Domains*
Data were collected from a domain with fast citation characteristics (a sub-domain of physics), and from a domain with slow citation characteristics (mathematics). In both domains, three adjacent partition cells are observed, containing all journals assigned to a particular combination of subject categories:
- 'Astronomy Astrophysics' only (A), 'Physics Particles Fields' only (P), and both 'Astronomy Astrophysics' and 'Physics Particles Fields' (A&P), with 8047, 1977 and 2440 articles respectively.
- 'Mathematics' only (M), 'Mathematics Applied' only (MA), and both 'Mathematics' and 'Mathematics Applied' (M&MA), with 10022, 3938 and 3286 articles respectively.

*Groups of highly cited publications*
Groups of highly cited publications were identified using both the technique of pre-set percentiles (5% and 1%), and the CSS-methodology (at least 'remarkably' and 'outstandingly' cited publications, stabilizing in size towards 8 to 11%, and 2 to 3% respectively in the largest citation windows).

**Results and discussion**

The citation distributions vary with domain and highly cited level. Peak citation years tend to fall later for more highly cited publications. Differences depend on the domain, with in the physics cells 0 to 3 years between the top-1% or outstandingly cited and the less cited publications, and in the mathematics cells 1 tot 2 years between the top-5% or at least remarkably cited and the less cited publications. In particular for publications below the top-1% and outstandingly cited, peak citation years in the slow mathematics domain fall later than in the fast physics domain (3 to 6, and 1 to 2 years after the publication year respectively). Figure 1 shows how groups of highly cited publications converge in content with increasing citation window length, in a similar way for both methodologies used. A yearly fluctuation in content remains between the larger consecutive citation windows.

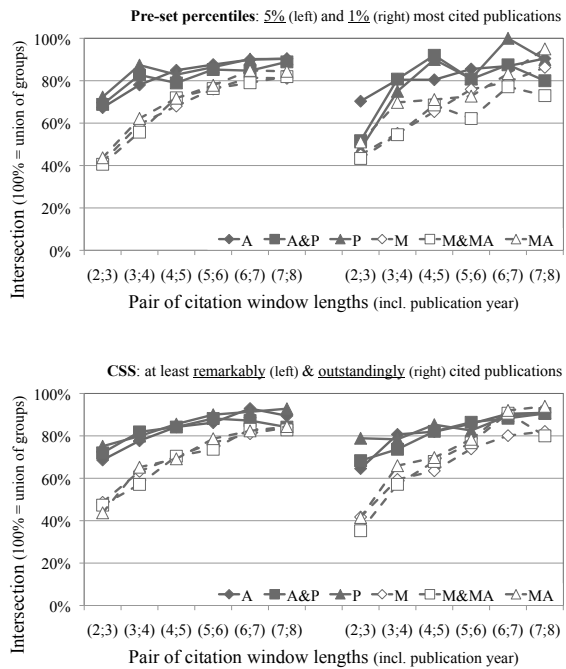

*Data sourced from Thomson Reuters Web of Knowledge (formerly referred to as ISI Web of Science). Web of Science (WoS) accessed online 10-11.01.2013 and 14.02.2013.*

**Figure 1. Intersection of groups of highly cited publications for consecutive citation window lengths.**

Convergence in content varies with domain in particular, and also with highly cited level. From a citation window of 3 years, groups of highly cited publications identified in consecutive citation windows have a majority of publications in common. This finding is in accordance with citation distribution models reflecting the conjecture that for most papers the initial pulse of citations determines its future citation history (Price, 1976). To have at least 80% of highly cited publications in common with the subsequent citation window requires 3 to 4 years for the physics cells, and 5 or more years for the mathematics cells. Changes in status for 1 out of 5 highly cited publications over the years might however still significantly influence a comparison of publication records of individual scientists. In practice citations are often collected in about the same period as the observed publications, for instance extended with one year. The results shown in Figure 1 indicate that in particular the potential error induced by a citation window as short as two years would need to be considered in the design of an approach for the identification of highly cited publications.

**Conclusions**

The results of the present study show how, in stable size groups of highly cited publications, the content of individual publications converges with an extending citation window. To the author's knowledge, no similar investigations of the extent to which individual publications enter and leave the group of highly cited publications as the citation window is enlarged, figure in scientific literature. The process evolves towards a limited remaining yearly fluctuation and depends on domain and highly cited level. Additional factors outside the scope of this paper may also exert an influence, such as the structure within which the highly cited publications are identified, including subject category as well as publication based structures. Stability in content of groups of highly cited publications is relevant in particular in a context of publication records of limited size, such as those of individual scientists. The studied domains with different bibliometric characteristics indicate that a same stability in content can require citation windows of different lengths depending on the domain. Whether a sufficient stability in content can be attained for successful general and comparative applications at the micro-level, with citation windows of acceptable size in an evaluative context, requires further investigation.

**Acknowledgments**

This paper is related to research on individual excellence carried out at the Flemish Centre for R&D Monitoring (ECOOM).